\documentclass[a4paper,10pt]{article}

\usepackage[english]{babel}
\usepackage{hyperref}
\usepackage{mathtools}
\usepackage{xspace}

\usepackage{framed}
\usepackage{tikz}
\usetikzlibrary{trees}
\usepackage{subcaption}
\usepackage{fullpage}
\usepackage{amsfonts}
\usepackage[utf8]{inputenc}
\usepackage{authblk}

\usepackage[hyperref=true,citestyle=authoryear,style=authoryear,citetracker=true,maxbibnames=10,maxcitenames=1,backend=bibtex,natbib=true]{biblatex}
\renewbibmacro{in:}{} 
\DeclareFieldFormat[article]{volume}{\mkbibbold{#1}} 
\DeclareFieldFormat[article]{title}{#1} 
\DeclareFieldFormat[article]{pages}{#1} 
\DeclareFieldFormat[article]{number}{\mkbibparens{#1}} 
\DeclareNameAlias{sortname}{last-first} 

\renewbibmacro*{journal+issuetitle}{%
  \usebibmacro{journal}%
  \setunit*{\addspace}%
  \printfield{volume}%
  \iffieldundef{number} 
     {} 
      {\printfield{number}}%
  \newunit}

\bibliography{VaRScaling}

\title{Value-at-Risk time scaling for long-term risk estimation\footnote{The views, thoughts and opinions expressed in this paper are those of the authors in their individual capacity and should not be attributed to UniCredit S.p.A. or to the authors as representatives or employees of UniCredit S.p.A.
The authors are grateful to the colleagues Valentina Cazzola, Matteo Formenti, Cinzia Fumagalli, Mattia Manzoni and Miriam Mazzone for helpful comments and suggestions on the topic.\newline \indent
This paper is currently being reviewed in \textit{The Journal of Risk}, \url{http://www.risk.net/type/journal/source/journal-of-risk}}}

\author[1,2]{Luca Spadafora\thanks{Corresponding Author, \href{mailto:luca.spadafora@dmf.unicatt.it}{luca.spadafora@dmf.unicatt.it}}}
\author[1]{Marco Dubrovich}
\author[1]{Marcello Terraneo}
\affil[1]{UniCredit S.p.A., Piazza Gae Aulenti 3, 20154 Milan, Italy}
\affil[2]{Faculty of Mathematical, Physical and Natural Sciences, Universit\`a Cattolica del Sacro Cuore,\newline Via dei Musei 41, 25121 Brescia, Italy}

\date{}

\begin{document}

\maketitle

\begin{center}
\vspace{-1cm}
\large(Pre-Print Version)
\end{center}

\begin{abstract}
In this paper we discuss a general methodology to compute the market risk measure over long time horizons and at extreme percentiles, which are the typical conditions needed for estimating Economic Capital.
The proposed approach extends the usual market-risk measure, ie, Value-at-Risk (VaR) at a short-term horizon and 99\% confidence level, by properly applying a scaling on the short-term Profit-and-Loss (P\&L) distribution.

Besides the standard square-root-of-time scaling, based on normality assumptions, we consider two leptokurtic probability density function classes for fitting empirical P\&L datasets and derive accurately their scaling behaviour in light of the Central Limit Theorem, interpreting time scaling as a convolution problem.
Our analyses result in a range of possible VaR-scaling approaches depending on the distribution providing the best fit to empirical data, the desired percentile level and the time horizon of the Economic Capital calculation.

After assessing the different approaches on a test equity trading portfolio, it emerges that the choice of the VaR-scaling approach can affect substantially the Economic Capital calculation.
In particular, the use of a convolution-based approach could lead to significantly larger risk measures (by up to a factor of four)  than those calculated using Normal assumptions on the P\&L distribution.

\end{abstract}

\section{Introduction}

Banks that received approval from the Supervisors to compute regulatory capital via Internal Model Methods (IMM) usually rely on short-horizon 99\% Value-at-Risk (VaR) models for the calculation of their own funds requirements.
The time horizon is usually motivated by the assumption that market positions can be hedged or liquidated in a short period\footnote{Such a framework is anyway under discussion within the Fundamental Review of the Trading Book (see \citet{FRTB})}.
Besides regulatory capital, computed according to the above-mentioned practices, banks have to provide, for Pillar II purposes, an estimate of the capital required to face losses with a longer time horizon (typically one year) and more conservative percentiles.
Such a cushion is usually referred to as Economic Capital (EC) for market risk, and it is aggregated to the other risks in order to assess internal capital requirements.
Its calculation is based on economic principles and is linked to the bank's own risk profile.
From a market risk management point of view, EC can be interpreted as a 1-year VaR of the bank's market exposure.

In principle, several approaches can be devised to compute losses at a 1-year time horizon,
for instance:
\begin{itemize}
\item Scenario generation (for the risk factors) and subsequent revaluation of the portfolio, obtaining a 1-year Profit and Loss (P\&L) distribution,
\item Extension of short-term market risk measures, computed for Pillar I purposes, to a longer time horizon and higher percentiles.
\end{itemize}
The first approach, although sound in principle, has some well-known drawbacks.
First, the direct calculation of such a long-term, high-percentile risk measure is undoubtedly challenging when using standard VaR models.
In fact, when employing historical simulation or Monte-Carlo (MC) methods, the accurate estimation of a 1-year P\&L distribution is not an easy task, either because of the limited length of available time series of risk-factor shocks or because of the questionable reliability of the assumptions on risk-factors dynamics needed in MC simulations.
Also, the need for determination of high percentiles of the distribution critically increases the number of required scenarios.
Moreover, scenarios for the risk factors strongly depend on hypotheses on the drift, which is not easy to estimate on a historical basis\footnote{Most VaR models rely on historical time-series to estimate their parameters.}
Finally, it embeds the assumption of \textit{freezing} the bank's positions during the entire time interval, while in reality the portfolio composition evolves in time due to hedging and rebalancing.

On the contrary, using the second methodology to apply a time scaling to the 1-day P\&L distribution one can bypass the above-mentioned difficulties.
In this case, it is assumed that hedging and rebalancing can be done (under liquidity assumptions) over the chosen time horizon, implying a constant level of risk taken by the bank in its trading activity.
Moreover, the second approach has the advantage of relying (for IMM banks) on models used into day-to-day activity and already approved to compute regulatory requirements.

In this paper we focus on the latter approach and we develop a robust methodology to extend Pillar I VaR to longer time horizons and extreme percentiles.
Although most of our considerations implicitly refer to VaR models based on historical simulation, the developed approach is valid also for MC models without loss of generality.
The only needed assumption is that the short time horizon at which VaR is computed can be safely assumed as a typical rebalancing time scale for the portfolio.
As a consequence, we model the P\&L over subsequent time steps using independent and identically distributed (iid) Random Variables (RVs).
The core of our approach lies in applying the scaling by means of convolution, and interpreting the results in the light of the Central Limit Theorem, specifically deriving conditions under which the long-term P\&L distribution converges to the Normal limit.
Our analyses result in a generalized VaR-scaling methodology to be used in the calculation of Economic Capital, depending primarily on the properties of the short-term P\&L distribution.
Essentially, using both analytical and empirical evidence, we show that:
\begin{itemize}
\item If the P\&L distribution has an exponential decay, the scaling of VaR can be correctly estimated using the standard square-root-of-time rule (SRTR), since the long-term P\&L distribution converges to the Normal quickly enough;
\item If the P\&L distribution has a power-law decay, the SRTR can be safely applied only if the tails are not too fat.
Otherwise, the SRTR cannot be applied and the long-term P\&L distribution has to be estimated explicitly.
In this latter case, the resulting EC can be significantly larger than what would be calculated assuming normality of the distribution.
\end{itemize}
While the specific results on EC calculations refer to confidence levels and time horizons of practical relevance, our analytical derivation is general.

The literature about VaR extrapolation with respect to quantile and time-horizon changes is quite large and diversified.
In general, most research papers deal with going beyond the standard Normal (SRTR) approaches, which are widespread among practitioners: there is an overall agreement on the inefficiency of the SRTR for the estimation of long-term VaR, since the underlying hypothesis of Normal RVs is usually not supported by empirical data.
This fact was already observed by \citet{Dorst} and \citet{Christoffersen} and further analysed in more recent years.
From this point of view, we deem this paper original in the sense that we found few references explicitly dealing with VaR extensions focusing on the portfolio P\&L distribution (rather than the RVs on which the P\&L distribution depends) and aiming at an accurate calculation of EC.
In fact, VaR scaling has been discussed with respect to the dynamics of the relevant RVs in several research papers, such as those of \citet{Kinateder} and, more recently, \citet{Oeuvray} and \citet{Degiannakis}.
Also the paper of \citet{Danielsson}, whose results will be recalled in more detail in the following, applies this approach to show that, when returns follow a stochastic jump-diffusion process, the use of the SRTR could induce a downward bias in risk estimation.
A similar effect about the bias dependence on the simulated process was observed also by \citet{Wang}.
In some sense, our work can be related with that of \citet{Skoglund}, where the scaling of VaR is analysed in connection with different trading strategies over time, although their work focuses on 1-day to 10-day extensions, not dealing with longer time horizons and extreme percentiles.
Further discussion on VaR time scaling, although not closely related to the results derived here, can be found in the works of \citet{Hamidieh} (extension of daily VaRs to 5 and 10 days using empirical scaling factors), \citet{Embrechts} (overview of long-term modelling methods for risk factors), \citet{Engle} (econometric model for the time behaviour of distribution quantiles) and in the references therein.

The paper is structured as follows.
In Section~\ref{VaR_Scaling} we introduce the concept of VaR scaling and the major challenges in its computation, besides reviewing the current market practices and their mathematical assumptions.
In Sections~\ref{PnL_Distributions}~and~\ref{Time_Scaling} we present our theoretical results on VaR scaling, starting from the assumptions on the underlying probability distributions.
We also provide some examples using empirical datasets to complement the highlights of the derivation.
Finally, in Sections~\ref{Application}~and~\ref{Summary_Conclusion} we apply our analytical results to a test equity trading portfolio and describe their implications in practical EC calculations.

\section{Value-at-Risk Scaling}\label{VaR_Scaling}

The problem of Value-at-Risk scaling concerns the extension of a certain VaR measure to a different confidence level or time horizon.
Formally, setting $t_0 = 0$, Value-at-Risk $\text{VaR}(\alpha,T)$ at confidence level $1 - \alpha$ and time horizon $T - t_0 = T$, is implicitly defined as:
\begin{equation}
1 - \alpha = \int_{-\infty}^{\text{VaR}} p(x(T))\mathrm{d}x(T)
\end{equation}
where $x(T)$ is the Profit \& Loss (P\&L) over time horizon $T$ and $p$ is its probability density function (PDF).
In general, the goal of VaR scaling is to estimate the unknown function $h(\cdot)$ such that:
\begin{equation}
\text{VaR}(\alpha',T') = h(\text{VaR}(\alpha,T))
\end{equation}
for $\alpha' \neq \alpha$ and $T' \neq T$.

In this work we assume, as is commonly done for practical reasons, a zero mean for both the short-term and the long-term P\&L distributions.
Over a 1-day horizon the empirical mean is usually very small, and can be assumed to be zero without any impact.
However this may not hold for the long-term distribution, since the mean propagates linearly with time (under the assumption of iid RVs), possibly growing to macroscopic values after the scaling.
This problem is discussed by \citet{Blake}, who show that using the SRTR and ignoring the mean in VaR calculations can lead to wrong estimations of the risk measure.
On the other hand, such a scaling of the mean would be difficult to justify in statistical terms, since the sample uncertainty of the empirical estimation of the short-term mean is typically comparable to the value itself.
Therefore we deem reasonable to set the long-term mean to zero, too, in order to ensure robustness and avoid dependence on initial conditions.

A common methodology to estimate the Economic Capital using VaR scaling 
is based on the assumption of normality of the P\&L distribution, and relies on the following steps:
\begin{itemize}
\item Application of the SRTR to compute the scaling of the chosen percentile
\begin{equation}\label{ECStd0}
x_{\alpha}(T') = x_{\alpha}(T)\cdot\sqrt{\frac{T'}{T}}
\end{equation}
with $x_{\alpha}$ denoting the percentile corresponding to confidence level $\alpha$;
\item Normal approximation of the P\&L distribution at any time-horizon (setting the mean to zero), therefore making the change of percentile a trivial task.
\end{itemize}
With this simple methodology, assuming that 1 year is equal to 250 (open) days, the Economic Capital at confidence level $1-\alpha$ can be calculated as:
\begin{equation}\label{ECStd}
\widehat{\text{VaR}}(\alpha,1y) = \sqrt{250} \frac{F_N^{-1}(\alpha)}{F_N^{-1}(0.01)} \widehat{\text{VaR}}(0.01,1d)
\end{equation}
where $F_N^{-1}(x)$ is the inverse Normal Cumulative Distribution Function (CDF) and $\widehat{\text{VaR}}(0.01,1d)$ is the estimated daily VaR at confidence level 99\%.

In this work we generalize this simplistic approach by deriving a VaR-scaling methodology based on the following steps:
\begin{enumerate}
\item Fit of the short-term (1-day) P\&L distribution, in order to choose the PDF with the highest explanatory power;
\item Calculation (either analytical or numerical) of the long-term (1-year) P\&L distribution, based on the chosen PDF class;
\item Computation of the Economic Capital as the desired extreme-percentile VaR-measure of the long-term P\&L distribution.
\end{enumerate}
Although our analytical results refer to a generic tail percentile $\alpha$, numerical examples are provided for confidence level $1 - \alpha^* = 99.93\%$\footnote{Historically a confidence level of 99.93\% was linked to the bank's target rating $A$ (according to the Standard \& Poor's scale) while, according to more recent surveys (see, eg, \citet{McK}), this relationship no longer holds.}.
Common choices for confidence levels in Economic Capital models typically range from 99.91\% to 99.99\%; as it will be shown in Section~\ref{Time_Scaling}, our results hold in this entire percentile range.

\section{Modelling P\&L distributions}\label{PnL_Distributions}

\subsection{Introducing theoretical PDFs}\label{Theoretical_PDFs}

The first step of our generalized VaR-scaling approach is to analyse empirical P\&L distributions in order to choose, among some candidates, the PDF with highest explanatory power.
In practice it is often difficult to justify which function class provides the best fit for empirical P\&Ls, since the amount of available data is usually not enough to span extreme percentiles efficiently\footnote{For example, in order to obtain on average at least one observation at 0.1\% percentile level one should consider about 1500 iid observations. A reasonable empirical estimation of that percentile level should require about $10^5$ observations. On the contrary, the typical number of available data is around $500$.}.
Consequently, we analyse fit performances also with respect to time setting and data availability.

The idea of this work is to benchmark the Normal distribution using two leptokurtic PDFs for fitting the datasets:

\begin{itemize}

\item Normal distribution (N)
\begin{equation}\label{Normal}
p_{N}(x;\mu,\sigma, T) = \frac{1}{\sqrt{2 \pi \sigma^2 T}}\exp{\left[\frac{(x - \mu T)^2}{2 \sigma^2 T}\right]}
\end{equation}
with daily mean $\mu$, daily standard deviation $\sigma$, and time horizon $T$.

\item Student's \textit{t}-distribution (ST)

\begin{equation}\label{t}
p_{ST}(x;\mu,\sigma,\nu) = \frac{\Gamma(\frac{\nu +1}{2})}{\sigma\sqrt{\nu\pi}\Gamma(\frac{\nu}{2})} \left[1 + \frac{(\frac{x - \mu}{\sigma})^2}{\nu}\right]^{-\frac{\nu+1}{2}}
\end{equation}
with mean $\mu$, scale factor $\sigma$ and $\nu$ degrees of freedom.
In this definition no time horizon $T$ appears, since the time scaling of the distribution is not available in a simple analytical form (this will be discussed below).
The Student's \textit{t}-distribution already has numerous successful applications in finance, ranging from the modelling of empirical equity returns (\citet{TS1}) to copula approaches for credit risk management (\citet{TS2}).

\item Variance-Gamma distribution (VG)
\begin{equation}\label{VG}
p_{VG}(x;\mu,\sigma,k,\theta, T) =  \frac{\sqrt{2}e^{\frac{\theta (x -\mu T)} {2\sigma^2}}}{\sigma\sqrt{\pi}k^{\frac{T}{k}}\Gamma(\frac{T}{k})}
\left( \frac{|x - \mu T|}{\sqrt{\frac{2\sigma^2}{k}+\theta^2}} \right)^{\frac{T}{k}-\frac{1}{2}} K_{\frac{T}{k}-\frac{1}{2}} \left(\frac{|x - \mu T| \sqrt{\frac{2\sigma^2}{k}+\theta^2}}{\sigma^2}\right)
\end{equation}
with daily mean $\mu$, scale factor $\sigma$, asymmetry factor $\theta$, shape factor $k$ and time horizon $T$.
The VG distribution (or VG process) was first introduced in finance to model asset returns~\citet{VG1}, and has subsequently been widely applied 
to option pricing~(see \citet{VG2}; \citet{VG3} and \citet{VG4}) and credit risk modelling~(\citet{VG5}).
\end{itemize}
In the above formulas $\Gamma(x)$ is the Gamma function and $K_{z}(x)$ is the modified Bessel function of the second kind.

These specific distributions were chosen as benchmark for the sake of convenience, besides being widely used by practitioners.
In fact, since the main problem of fitting returns (or P\&Ls) lies in the determination of how \textit{fat} are the tails of the distribution, we chose ST and VG because they have different asymptotic behaviours, while being leptokurtotic.
More specifically, VG decays exponentially whereas ST behaves as a power law:

\begin{equation}\label{tLimit}
p_{ST}(x) \approx \frac{\Gamma(\frac{\nu +1}{2})}{\sqrt{\pi}\Gamma(\frac{\nu}{2})} \frac{\left(\sigma \sqrt{\nu}\right)^{\nu}}{x^{\nu + 1}}
\end{equation}
\begin{equation}
p_{VG}(x) \sim \exp\left[- \left(\frac{\sqrt{\theta^2 + \sigma^2/k}}{\sigma^2} \pm \frac{\theta}{\sigma^2}\right) |x| \right]
\end{equation}
As a consequence of these different trends, risk estimations can be strongly affected by the choice of the underlying PDF and the desired confidence level.

\begin{figure}[t]
\centering
\begin{subfigure}[b]{0.475\textwidth}
\includegraphics[width=\textwidth]{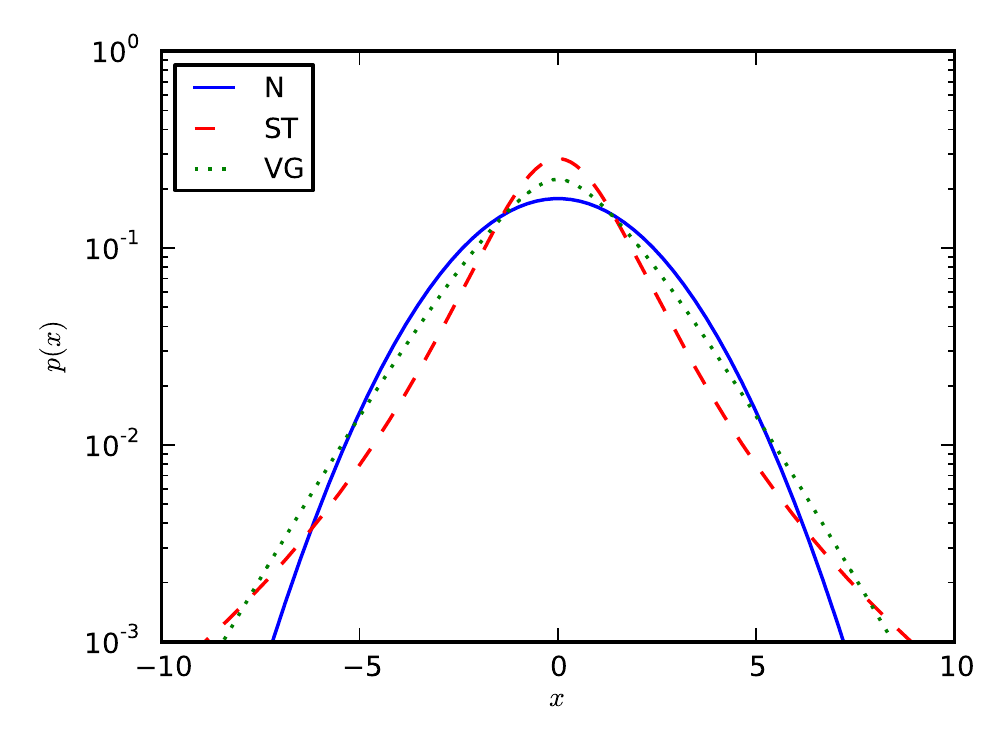}
\caption{PDF}
\label{PDF_Comparison}
\end{subfigure}
\begin{subfigure}[b]{0.475\textwidth}
\includegraphics[width=\textwidth]{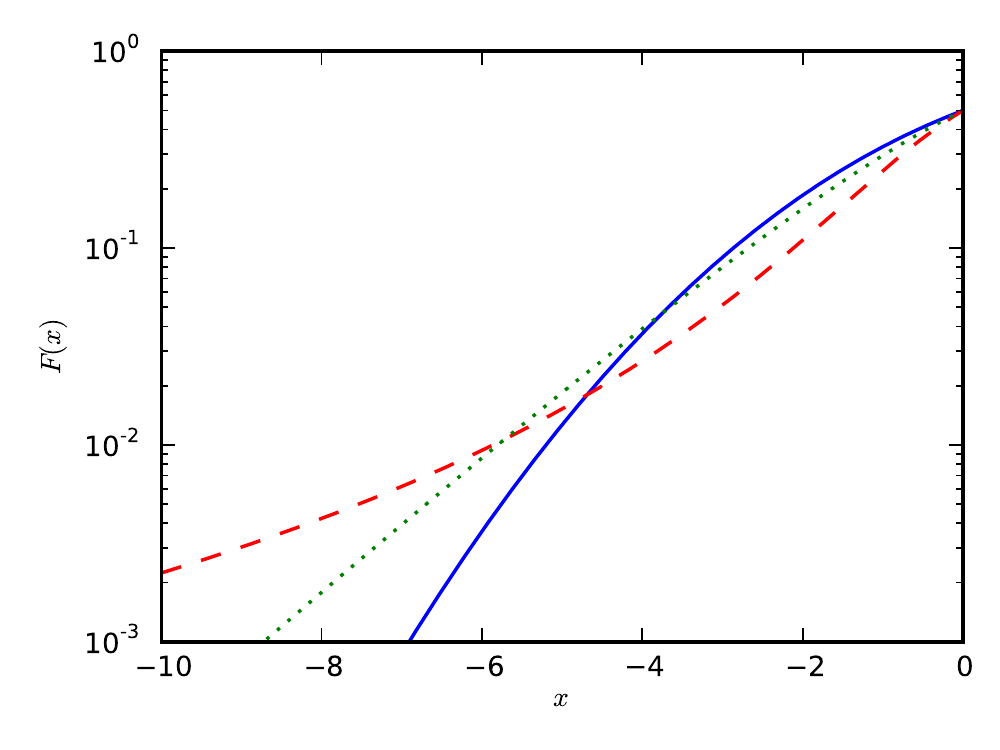}
\caption{CDF}
\label{CDF_Comparison}
\end{subfigure}
\caption{Comparison of Normal (N), Student's \textit{t}- (ST) and Variance-Gamma (VG) distributions (semi-log scale) on the whole real axis (PDF, left) and in a percentile region of interest (CDF, right). The ST degrees of freedom are $\nu = 3$ and the VG asymmetry and shape factors are $\theta = 0$ and $k = \frac{1}{2}$, respectively. All distributions have mean $\mu = 0$ and variance $\sigma^2 = 5$.}
\label{PDF_Comparison+CDF_Comparison}
\end{figure}

For a visual comparison, in Fig.~\ref{PDF_Comparison+CDF_Comparison} we plot the three distributions with the same mean and variance and attempt to estimate VaR at different confidence levels.
The probability that an extreme event occurs depends on the confidence level used to define the extreme event \textit{itself}.
For the confidence level range $0.011 < \alpha < 0.047$ (corresponding to percentiles $-5.718 < x_\alpha < -3.750$) VG implies a VaR larger than both Normal and ST distributions while, at lower confidence levels $\alpha < 0.011$ (or $x_\alpha < -5.718$) the largest VaR is implied by ST, because of its slower power-law decay.
The Normal CDF implies the largest VaR only at high confidence levels ($\alpha > 0.047$ or $x_\alpha > -3.750$).

\subsection{Fit performances over time}\label{Fit_Performances}

We tested the performance of the three considered function classes in explaining the 1-day P\&L distributions of the IBM stock over 1 year.
In all the cases, to match practical conditions, the number of observations (ie, the historical depth of the considered 1-day P\&Ls) is $N = 500$\footnote{This means using roughly two years of data for populating each P\&L distribution, a typical dataset for historical VaR calculations.}.
The fit was obtained by minimizing the mean squared error (MSE) between theoretical and observed CDFs in the cases of ST and VG distributions, and by moment matching in the Normal case.
In general, the observed P\&L distribution is hardly compatible with a Normal while, on the contrary, VG and ST give better fit results.

\begin{figure}[h]
\begin{center}
\includegraphics[width=\textwidth]{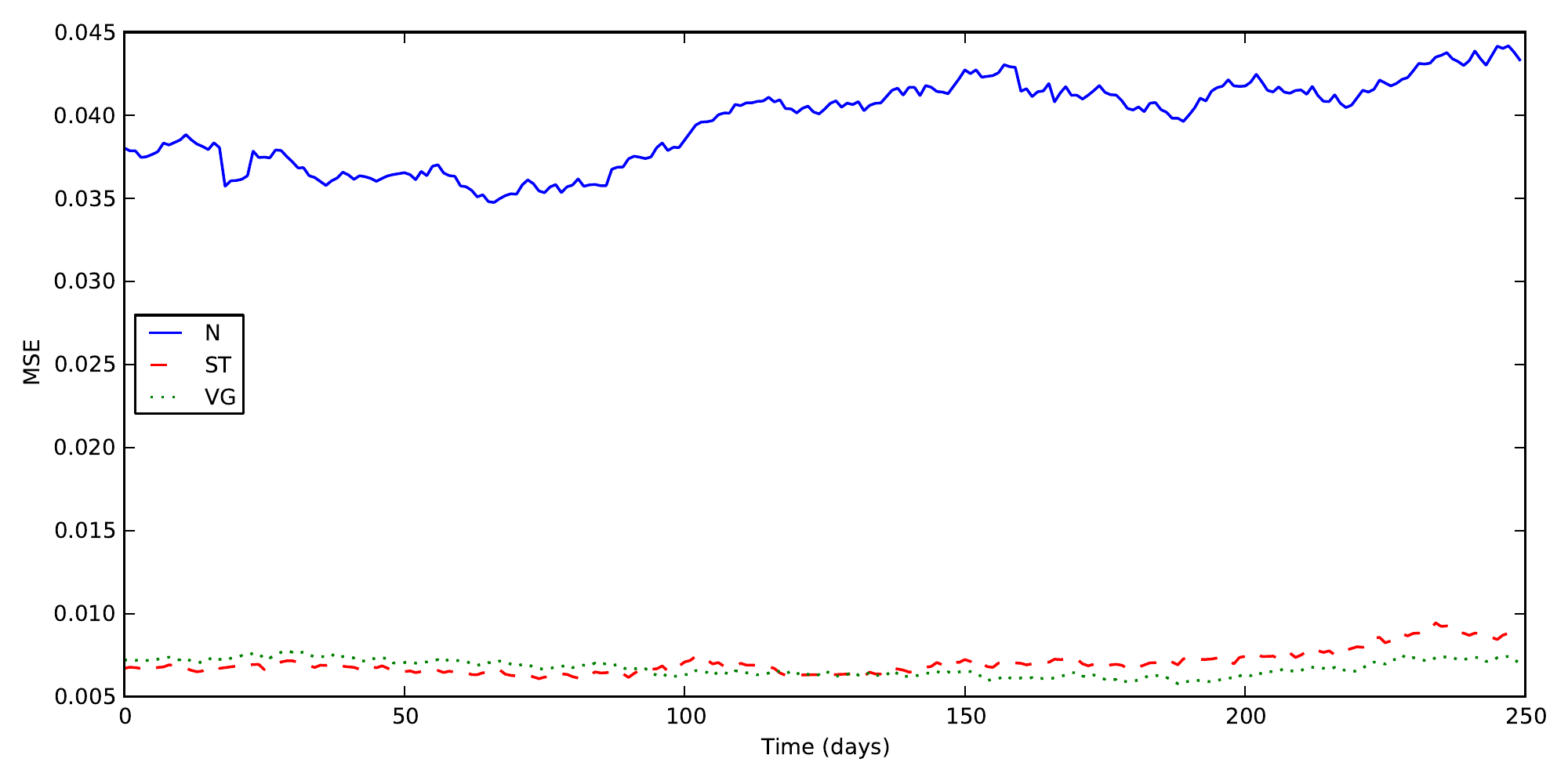}
\end{center}
\caption{Interpolation error over time for the Normal (N), Student's \textit{t}- and Variance-Gamma (VG) distributions obtained considering 250 P\&Ls strips (each made by 500 observations) of IBM. Data refers to the period 15/09/2011-12/09/2012.}
\label{MSE_Trend}
\end{figure}

\begin{figure}[h]
\centering
\begin{subfigure}[b]{0.475\textwidth}
\includegraphics[width=\textwidth]{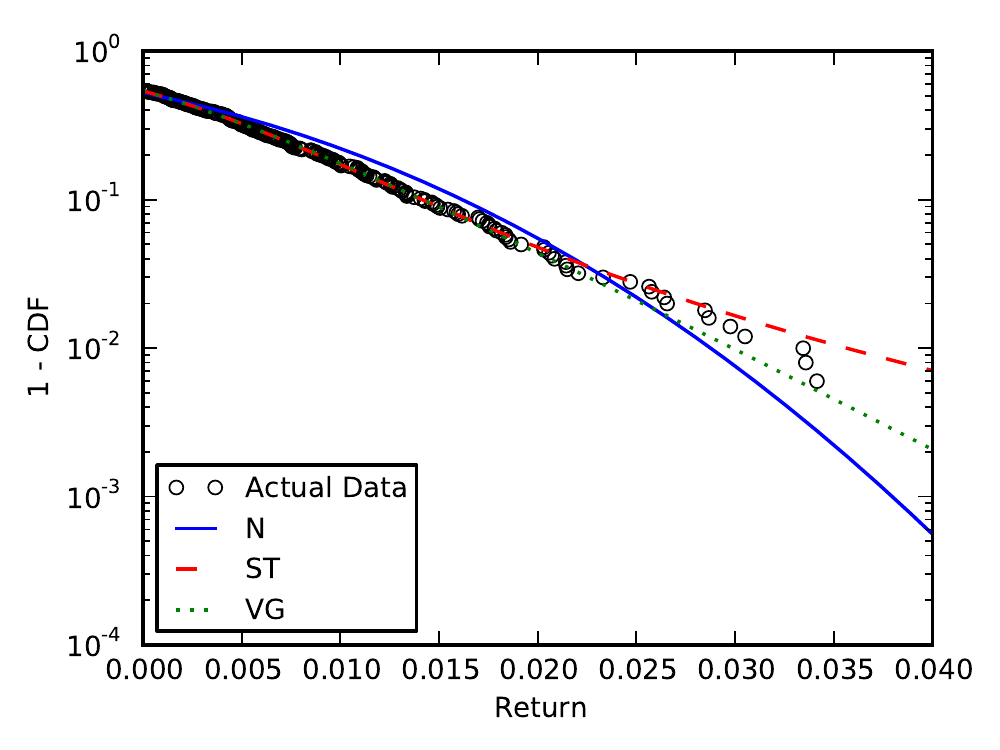}
\caption{Date:  21/09/2010.}
\label{IBM_CDF_Fit_ST}
\end{subfigure}
\begin{subfigure}[b]{0.475\textwidth}
\includegraphics[width=\textwidth]{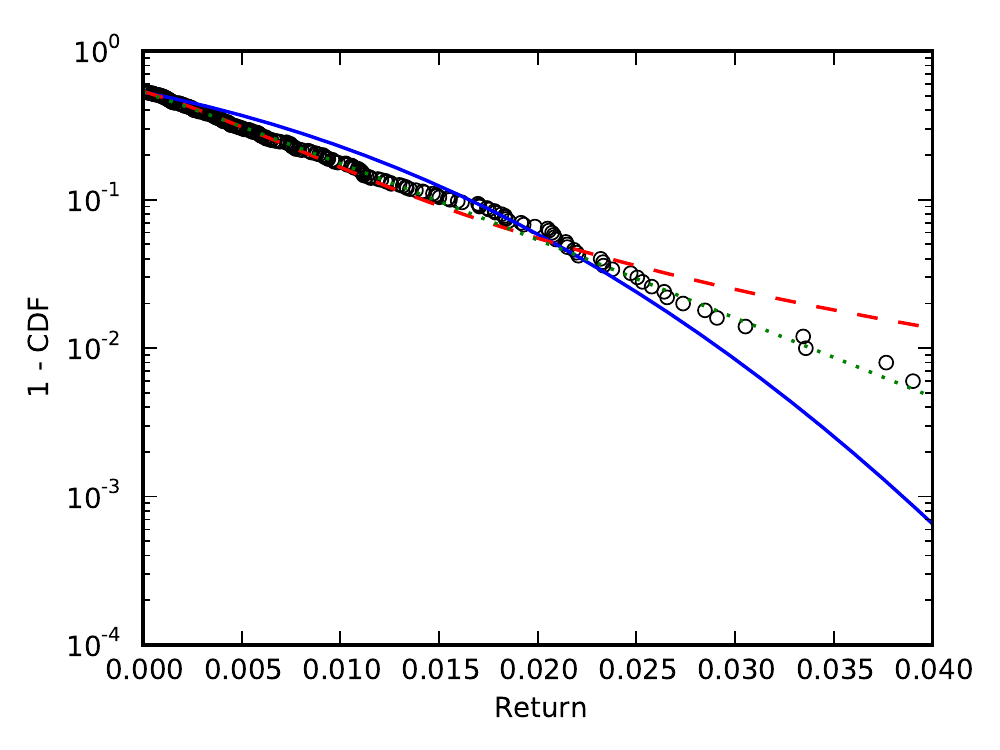}
\caption{Date: 12/09/2012.}
\label{IBM_CDF_Fit_VG}
\end{subfigure}
\caption{Fit of IBM's empirical CDF of returns with Normal (N), Student's \textit{t}- (ST) and Variance-Gamma (VG) distributions. While on 21/09/2010 (left) the interpolation error for the ST is smaller than for VG, the outcome is opposite on 12/09/2012 (right).}
\label{IBM_CDF_Fit_ST+IBM_CDF_Fit_VG}
\end{figure}

In Fig.~\ref{MSE_Trend} we show the MSE trend for each theoretical CDF using 250 P\&L strips obtained considering a rolling window of 500 observations: in general it is difficult to determine whether the VG or ST distribution provides a better agreement with empirical data.
Consider, for example, the IBM P\&L distribution as of dates 21/09/2010 and 12/09/2012, which we plot in Figs.~\ref{IBM_CDF_Fit_ST}~and~\ref{IBM_CDF_Fit_VG}, respectively.
In the former case the better fit is achieved by ST (smaller MSE than VG), while in the latter the outcome is opposite.


\subsection{Data availability}\label{Data_Availability}

\begin{figure}[h]
\centering
\begin{subfigure}[b]{0.475\textwidth}
\includegraphics[width=\textwidth]{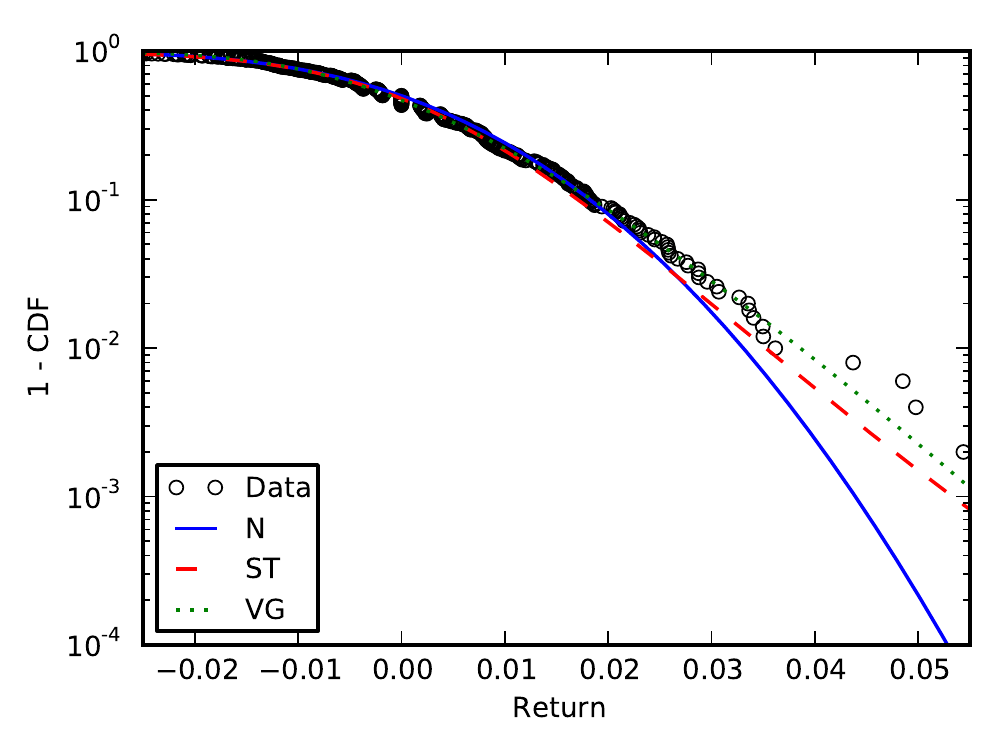}
\caption{Number of observations: 500}
\label{IBM_CDF_Fit_Short}
\end{subfigure}
\begin{subfigure}[b]{0.475\textwidth}
\includegraphics[width=\textwidth]{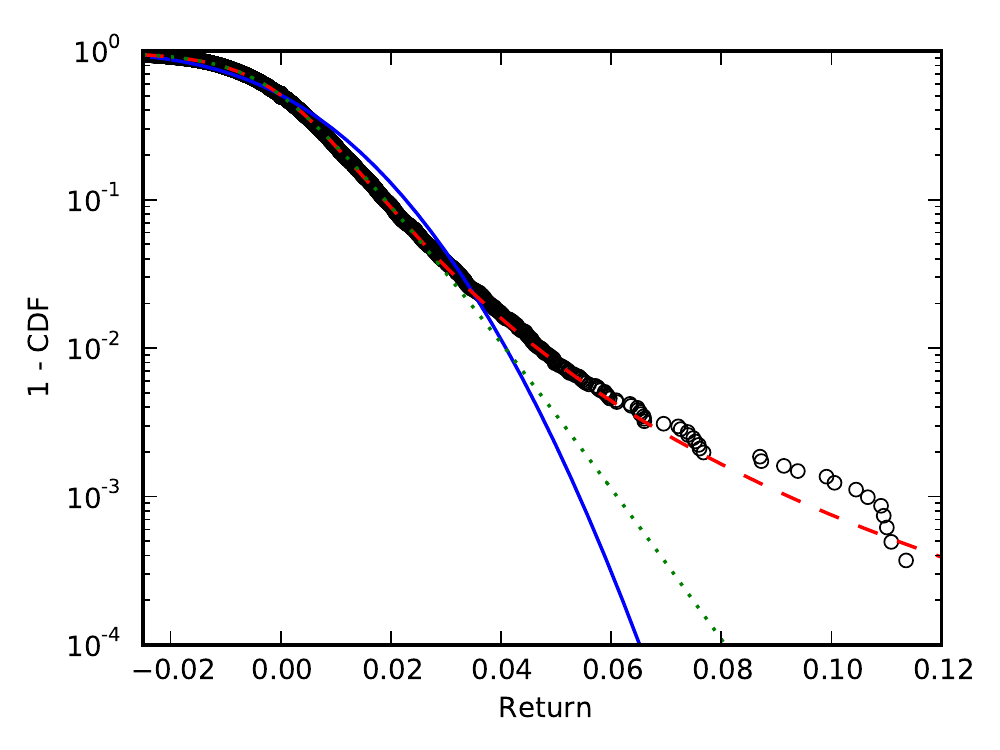}
\caption{Number of observations: 8000}
\label{IBM_CDF_Fit_Long}
\end{subfigure} 
\caption{Fit of IBM's empirical CDF of returns with Normal (N), Student's \textit{t}- (ST) and Variance-Gamma (VG) distributions, considering a subset of the historical time series (left) and the whole available time series (right). While with a small number of observations VG and ST yield similar fit results, with 8000 observations ST clearly provides the best performance.}
\label{IBM_CDF_Fit_Short+IBM_CDF_Fit_Long}
\end{figure}

Uncertainty on the choice of the theoretical CDF yielding better fit results is often increased by the limited sample size.
As an example, we focused again on the fit performances for IBM's returns.
When using an arbitrary subset of 500 observations the fit results do not allow to state whether VG or ST provides a better description of the tails (Fig.~\ref{IBM_CDF_Fit_Short}) (see above discussion).
On the contrary, when considering the whole available time series (8000 observations) the outcome changes.
In this case it is evident that the ST distribution has a better explanatory power for the tail behaviour of the distribution (Fig.~\ref{IBM_CDF_Fit_Long}).

In practice, the number of observations available for VaR measurements at portfolio-level is typically of the order of 500, meaning that the tail behaviour of the P\&L distribution cannot be empirically characterized with certainty, and further analyses are required.
\\\\
In conclusion, the determination of the PDF providing the best fit to empirical P\&L distributions is a challenging task.
Within a VaR-scaling framework this is even more crucial as it poses the basis on which subsequent numerical or analytical calculations are carried out.
In general, any assumption on the theoretical CDF to be used for economic capital calculations shall be justified with great care after in-depth analyses, as its empirical characterization is in principle not possible.
One may say that these difficulties compensate the relatively straightforward methodology that lies behind scaling techniques for extending market-risk measures.

\section{Time Scaling}\label{Time_Scaling}

\subsection{Computing the long-term P\&L distribution: Convolution and the\\ Central Limit Theorem}\label{Convolution_CLT}

As introduced in Sec.~\ref{VaR_Scaling}, in practice it is very difficult to obtain P\&L distributions over long time horizons.
This problem can be overcome by borrowing some concepts from functional analysis: the long-term PDF can be calculated (either analytically or numerically) starting from the short-term PDF by means of \textit{convolution}.

Suppose we start at time zero with the knowledge of the P\&L distribution over time horizon $\Delta t$ and need to obtain the P\&L distribution over time horizon $T = n \Delta t$.
Formally, we denote with $X(\Delta t)$ the RV with values $x(\Delta t) \in \mathbb{R}$ representing the P\&L over time horizon $\Delta t$.
At each time-step $\left\lbrace \Delta t, 2 \Delta t, ..., n \Delta  t = T\right\rbrace$ a new draw is performed from the same distribution.
Denoting the P\&L from time $(k-1) \Delta  t$ to time $k \Delta  t$ with $x_k(\Delta t)$, the realized P\&L from time zero to time horizons $\left\lbrace \Delta t, 2 \Delta t, ..., n \Delta t = T\right\rbrace$ is given by the sequence
\begin{align}
\text{P\&L}_{0 \rightarrow \Delta t} &= x_1(\Delta t)\nonumber\\
\text{P\&L}_{0 \rightarrow 2 \Delta t} &= \text{P\&L}_{0 \rightarrow \Delta t} + x_2(\Delta t) = x_1(\Delta t) + x_2(\Delta t)\nonumber\\
&\vdots\nonumber\\
\text{P\&L}_{0 \rightarrow T} &= \text{P\&L}_{0 \rightarrow (n-1) \Delta t} + x_n(\Delta t) = x_1(\Delta t) + x_2(\Delta t) + ... + x_n(\Delta t)\nonumber\\
&= \sum_{k=1}^{n} x_k(\Delta t)
\end{align}
On the other hand, the PDF of the sum of two independent RVs is given by the convolution of their individual distributions.
In our case this reads
\begin{equation}
p(y) = \int_{-\infty}^{+\infty} p(y - x_1(\Delta t))p(x_1(\Delta t))\mathrm{d}x_1(\Delta t)
\end{equation}
with $y = x_1(\Delta t) + x_2(\Delta t)$, for the first step ($\Delta t \rightarrow 2 \Delta t$) and continues similarly up to $T = n \Delta t$.
Applying $n$ times the convolution operator to the initial $\Delta t$-P\&L distribution it is possible to obtain the $T$-P\&L distribution, which is the one that is relevant to Economic Capital calculations.
In the Normal case, this result leads to well-known SRTR.
An analytical form for the convolution can be obtained also for the VG distribution (see Eq.~\eqref{VG}) while we rely on the convolution theorem combined with the Fast Fourier Transform (FFT) algorithm to estimate the convolution in the ST case.

At this point, a natural question would be if it is possible to obtain an asymptotic behaviour of the $n$-times ($n \rightarrow \infty$) convoluted PDF for a given distribution family.
As we are dealing with a \textit{sum} of iid RVs it is possible to apply the Central Limit Theorem (CLT)\footnote{We use here the classical Lindeberg-L\'{e}vy formulation.} to our problem.
Specifically, if $\mu \Delta t$ and $\sigma \sqrt{\Delta t}$ are the expected value and the standard deviation of RV $X$, under some mild assumptions on its PDF $p_D(x;\cdot)$ the $n$-times convoluted distribution of $X$ satisfies:
\begin{equation}
\lim_{n\rightarrow +\infty} \left\lbrace P(\alpha <\sum_{i=1}^n x_i< \beta)\right\rbrace = \int_{\alpha}^{\beta} \frac{1}{\sqrt{2\pi\sigma^2 n \Delta t}}e^{-\frac{(x - \mu n \Delta t)^2}{2\sigma^2 n \Delta t}} 
\end{equation}
for all \textit{finite} $\alpha$ and $\beta$.
In simple words, the distribution of the sum of iid RVs converges (as $n$ goes to infinity) to a Normal distribution with a suitably rescaled mean and variance.
Formally, the above identity holds only when $n \rightarrow \infty$; for finite $n$ it is understood that the CLT only concerns the central region of the distribution, which has to be quantified in some way.

This crucial result implies that, if the percentile considered for VaR estimation purposes falls into the central region of the distribution in the sense of the CLT after convolving, it will be possible to approximate the $n$-times convoluted distribution with a Normal.
For example, if we assume that at time $T = n \Delta t$ the CLT holds for percentile $x_\alpha$ of a generic distribution $p_D(x;\cdot)$, then
\begin{equation}\label{CLTperc}
VaR_{D}(\alpha,T) \simeq F_N^{-1}(\alpha;\mu T,\sigma \sqrt{T}) = VaR_{N}(\alpha,T)
\end{equation}
where $F_N^{-1}(x; \mu T,\sigma \sqrt{T})$ is the Normal inverse CDF with mean $\mu T$ and standard deviation $\sigma \sqrt{T}$.
When, on the contrary, the number of time steps is not enough to obtain convergence, the resulting P\&L distribution has to be computed by convolving the initial PDF $n$ times.
This approach is more problematic since its outcome depends strongly on the quality of the initial PDF characterization, besides being computationally intensive.

\begin{figure}[t]
\begin{center}
\includegraphics[width=0.5\textwidth]{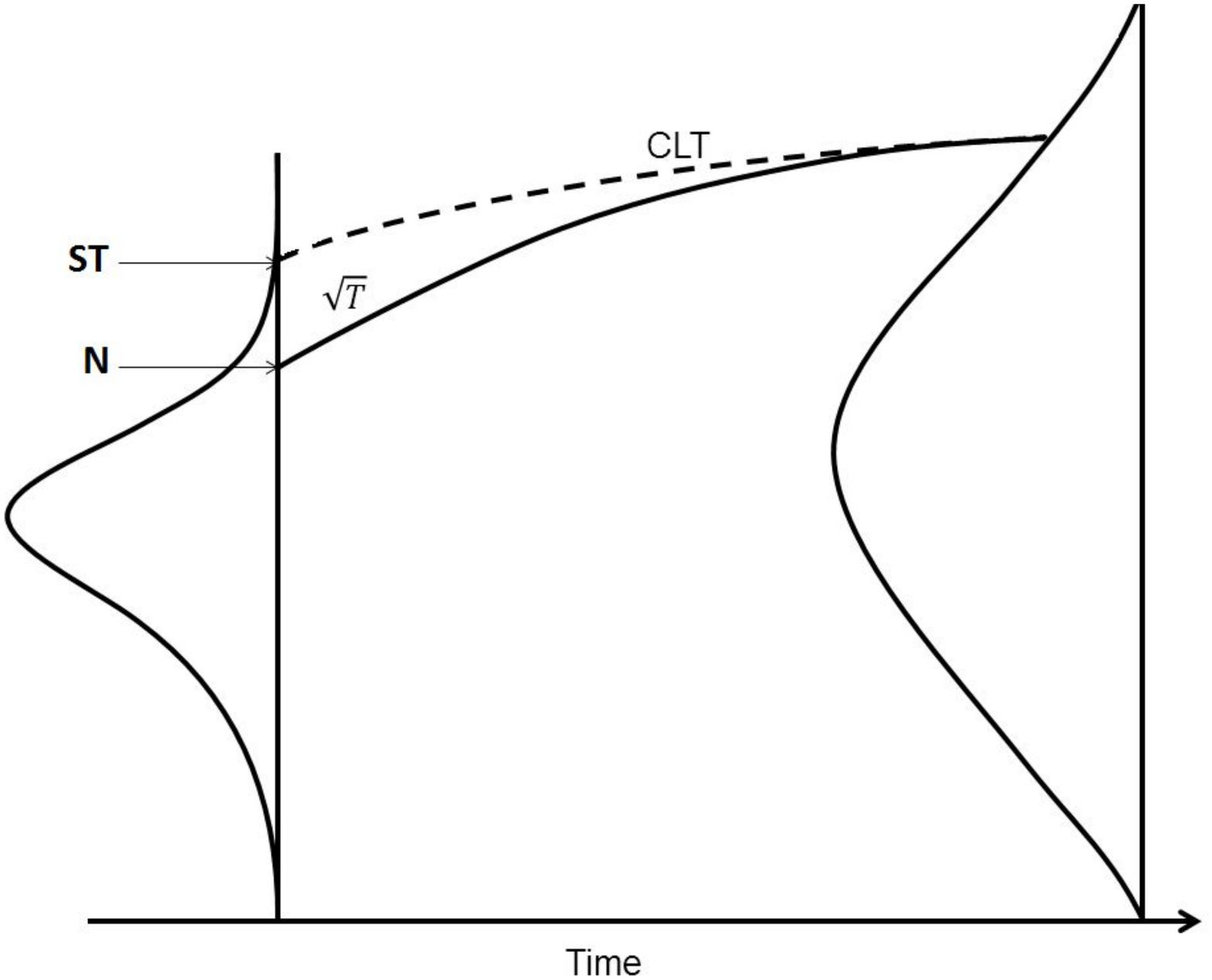}
\end{center}
\caption{Schematic representation of the time scaling of Normal (N) and Student's \textit{t}- (ST) distributions. If the CLT holds (ie, if $T$ is large enough), the ST VaR converges to the Normal VaR.}
\label{VaR_Evolution}
\end{figure}

The above considerations have a deep meaning.
Take, for example, the ST distribution in a percentile region such that, for $\Delta t=1d$, $VaR_{ST}(\alpha,1d) > VaR_{N}(\alpha,1d)$.
Then, from Eq.~\eqref{CLTperc} it is clear that, by continuity assumptions, the time scaling for ST should be slower than $\sqrt{T}$, as sketched in Fig.~\ref{VaR_Evolution}.
From this point of view, a naive SRTR approach focused on the scaling of a certain extreme percentile (as in Eq.~\eqref{ECStd0}) can be deemed conservative, overestimating the measures of risk:
$$VaR_{ST}(\alpha, 1d) \sqrt{T} > VaR_{ST} (\alpha,T)\geq VaR_{N}(\alpha,T)$$
However, put in another way, the above result implies that, whenever the percentile considered for VaR estimation does not fall into the CLT convergence region, the VaR implied by ST is always larger than the standard Normal VaR.
These results can be considered a generalization of those of \citet{Danielsson}, since their derivation relies on the assumption that the CLT holds and this is not the case for the very large percentiles considered here.
The bias in the risk estimation, consequently, can be positive or negative depending on the confidence level considered for VaR estimations.

\subsection{Convergence to the Normal limit}\label{Normal_Convergence}

\begin{figure}[t]
\begin{center}
\begin{subfigure}[b]{\textwidth}
\includegraphics[width=\textwidth]{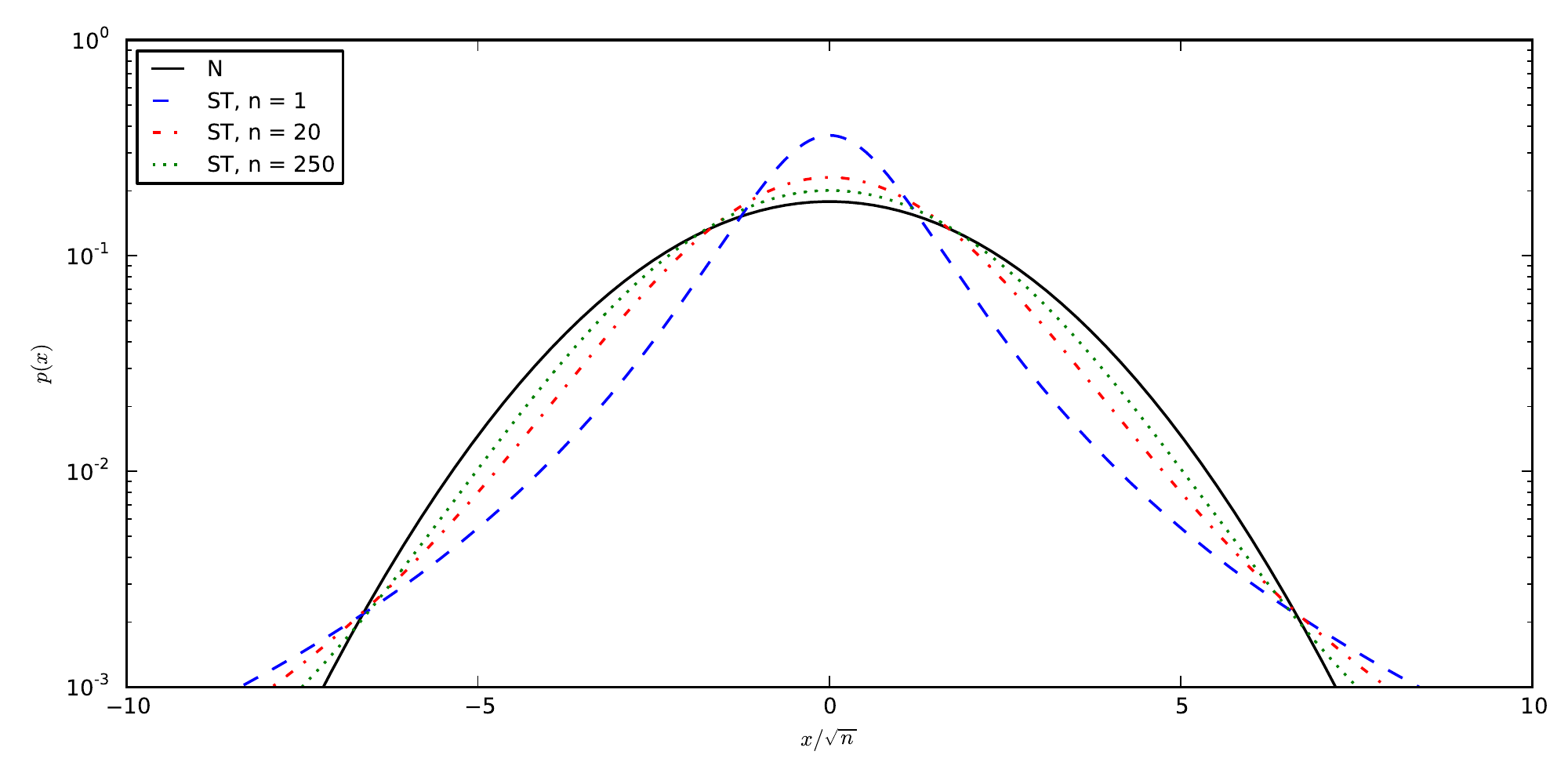}
\caption{Student's \textit{t}-distribution.}
\label{ST_Convergence}
\end{subfigure}\\
\begin{subfigure}[b]{\textwidth}
\includegraphics[width=\textwidth]{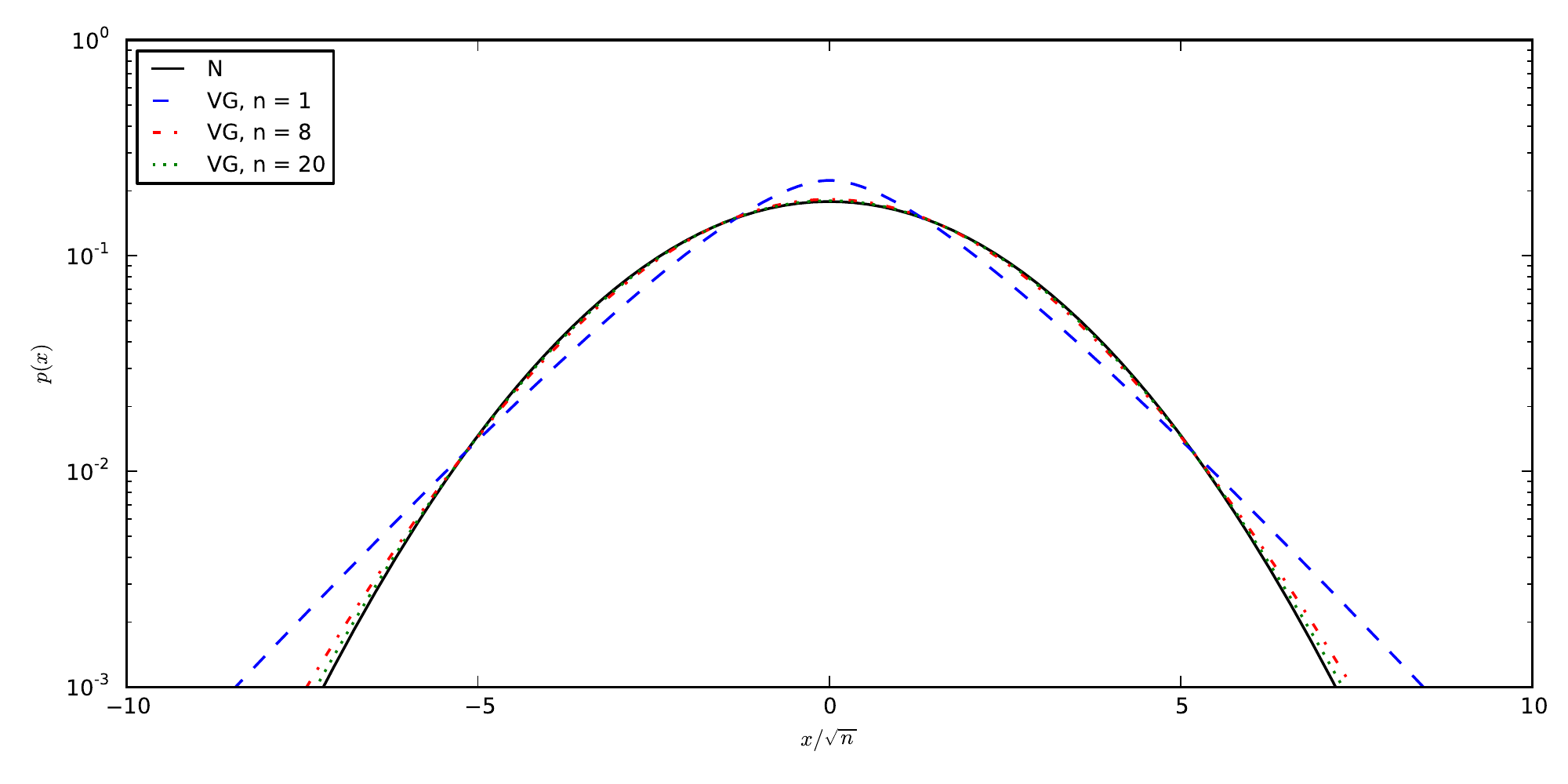}
\caption{Variance-Gamma distribution.}
\label{VG_Convergence}
\end{subfigure}
\end{center}
\caption{Convoluted Student's \textit{t}- (top) and Variance-Gamma (bottom) distributions at the increase of time, on top of the limit-case Normal distribution. In the ST case, the differences between the two are still visible after $n = 250$ convolution steps. The starting variance is $\sigma^2 = 5$ in all cases. The other parameters are the same as those of Fig.~\ref{PDF_Comparison+CDF_Comparison}.}
\label{ST_Convergence+VG_Convergence}
\end{figure}

The conditions under which the central region of a given distribution is sufficiently ``Normal" have to be quantified.
Moreover, the central region itself needs to be defined in some sense.
For a visual comparison, in Fig.~\ref{ST_Convergence+VG_Convergence} we plot a number of subsequent convolutions of ST and VG PDFs (with the same initial variance) besides the limit-case Normal distribution.
It is already clear from the plots that, while in the VG case convergence is achieved after few convolutions, in the ST case the trend is much slower.
We analyse both cases in the following.

\subsubsection{Student's \textit{t}-distribution}\label{ST_Normal_Convergence}

We propose here a method for determining when, and for which value of parameter $\nu$, the convoluted ST distribution converges (up to the desired quantile $\alpha$) to the Normal distribution.
We are interested in defining a critical P\&L value $x^{*}$ beyond which the two distributions start to become substantially different.
Following the derivation of~\citet{Bouchaud}, we define this quantity implicitly as the point where the two distributions intersect before decaying to zero at infinity.
Using Eqs.~\eqref{Normal}~and~\eqref{tLimit} we have:

\begin{equation}\label{EqCLT}
\frac{1}{\sqrt{2\pi \sigma^2 \left( \frac{\nu}{\nu-2} \right) T}}
 \exp \left[-\frac{x^2}{2 \sigma^2 \left( \frac{\nu}{\nu-2} \right)T } \right] = \frac{ \Gamma \left( \frac{\nu +1}{2}\right) }{\sqrt{\pi} \Gamma \left( \frac{\nu}{2} \right) } \frac{ \left( \sigma\sqrt{\nu} \right)^\nu T}{x^{\nu + 1}}
\end{equation}
where the variance of the ST distribution is given by $\sigma^2\nu/(\nu-2)$.
To solve Eq.~\eqref{EqCLT}, we guess a solution of the form

\begin{equation}\label{ansatz}
x^* = \sigma \sqrt{\frac{\nu}{\nu-2}}\sqrt{T \mathrm{log} (T^{\psi})}
\end{equation}
where $\psi$ is a parameter to be adjusted.
Substituting Eq.~\eqref{ansatz} in Eq.~\eqref{EqCLT} we have

\begin{equation}\label{x_star_intersection}
\frac{1}{\sqrt{2\pi}T^{\frac{\psi +1}{2}}} = \frac{ \Gamma \left( \frac{\nu +1}{2}\right) }{\sqrt{\pi} \Gamma \left( \frac{\nu}{2} \right) } \frac{ (\nu -2)^{\nu/2}}{T^{\frac{\nu-1}{2}}\left(\mathrm{log}(T^{\psi})\right)^{\frac{\nu + 1}{2}}}
\end{equation}
The above equation cannot be solved exactly for $\psi$.
Since in this context we are in any case interested in deriving an analytical approximation for the quantity $x^*$, we now simplify Eq.~\eqref{x_star_intersection} taking into account just the leading terms.
In order to match the convergence speed as $T$ increases we have:

\begin{equation}
\frac{\psi + 1}{2} = \frac{\nu - 1}{2}
\end{equation}
or $\psi = \nu -2$.
Substituting this result in Eq.~\eqref{ansatz} we obtain the time-scaling behaviour of the critical value

\begin{equation}\label{xCritical}
x^* = \sigma \sqrt{\nu} \sqrt{T \mathrm{log}(T)}
\end{equation}
This result shows that, as already visible in Fig.~\ref{ST_Convergence}, the convergence to the Normal is relatively slow as time increases.

\begin{figure}[h]
  \begin{center}
    \includegraphics[width=0.5\textwidth]{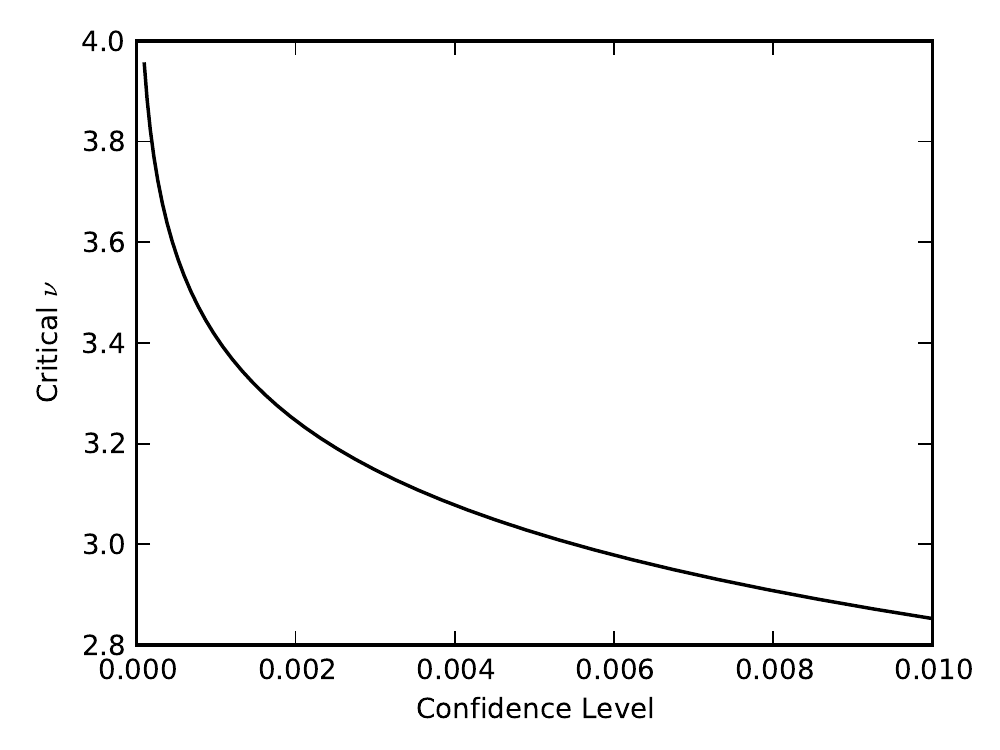}
  \end{center}
  \caption{Critical value of $\nu$ as a function of the tail confidence level, calculated according to Eq.~\eqref{Pcritical}.}
  \label{Nu_Critical}
\end{figure}

Starting from Eq.~\eqref{xCritical}, it is possible to estimate the percentile level at which the convergence condition for the ST distribution (defined above) is satisfied after exactly one year, as a function of $\nu$\footnote{Note that this relationship does not depend on the parameter $\sigma$.}:

\begin{align}\label{Pcritical}
P(\sigma\sqrt{\nu}\sqrt{T \mathrm{log}(T)} < x < +\infty) &= \int\limits_{\sigma\sqrt{\nu}\sqrt{T \mathrm{log}(T)}}^{+\infty} \frac{ \Gamma \left( \frac{\nu +1}{2}\right) }{\sqrt{\pi} \Gamma \left( \frac{\nu}{2} \right) } \frac{\left(\sigma\sqrt{\nu}\right)^\nu T}{x^{\nu +1}} \mathrm{d}x\\
&= \frac{(\nu +1)\Gamma\left(\frac{\nu + 1}{2}\right)}{\sqrt{\pi}\Gamma\left( \frac{\nu}{2} \right)} \frac{1}{T^{\frac{\nu -2}{2}}\left(\mathrm{log}(T)\right)^{\nu/2}}
\end{align}
If we impose in the above formulas $P = 0.07\%$ and $T = \text{1 year}$, we obtain $\nu^* = 3.41$.
This means that, using the above-defined criterion for defining the CLT convergence region for the ST distribution, the critical value $\nu^*$ discriminates between
\begin{itemize}
\item Convergence regime ($\nu > \nu^*$): for large values of $\nu$ convergence is achieved before $n = 250$ convolution steps.
This agrees with the limit condition ($\nu \rightarrow +\infty$) for which the ST distribution becomes a Normal distribution
\item Non-convergence regime ($\nu < \nu^*$): for small $\nu$, even after $n = 250$ convolutions the PDF is significantly different from its Normal limit.
Indeed, the fewer the degrees of freedom the fatter the tails of the ST, and the more different from the Normal
\end{itemize}
In Fig.~\ref{Nu_Critical} we show the dependence of the critical value $\nu^*$ on the percentile considered for the economic capital calculation.
The qualitative outcomes of our analysis hold for all the percentile levels typically considered in such measures.
As expected, as the desired confidence level becomes smaller, the implied critical value $\nu^*$ becomes larger, meaning a slower and slower convergence of the convoluted distribution to the Normal.

We stress that the convergence condition derived above (Eqs.~\eqref{EqCLT}~to~\eqref{Pcritical}) relies on an asymptotic approximation of the leading terms in the expressions.
Therefore, it is meant to be used as a rule-of-thumb for discerning between different limit cases.
We will show in Section~\ref{Application} and Appendix~\ref{Appendix_A} that essentially our $\nu^*$ efficiently discerns between convergence and non-convergence regimes.
In particular, the relative difference between EC estimated using the Normal distribution and EC estimated using the ST distribution is small when $\nu\geq\nu^*$.

\subsubsection{Variance-Gamma distribution}\label{VG_Normal_Convergence}

In the Variance-Gamma case the convergence to the Normal distribution takes place in a much quicker way, in principle because both distributions are exponentially decaying.
Instead of defining an analytical criterion for checking convergence to the Normal (as done in the ST case above), we provide just a numerical example to prove that convergence is already achieved at extreme percentiles even after few convolutions.
Since in a typical VaR-scaling exercise one would have to extend a daily or a weekly VaR measure to an observation time of 1 year, it is sufficient to show that for $n \geq 50$ the two CDFs are hardly distinguishable.

\begin{figure}[h]
  \begin{center}
    \includegraphics[width=0.5\textwidth]{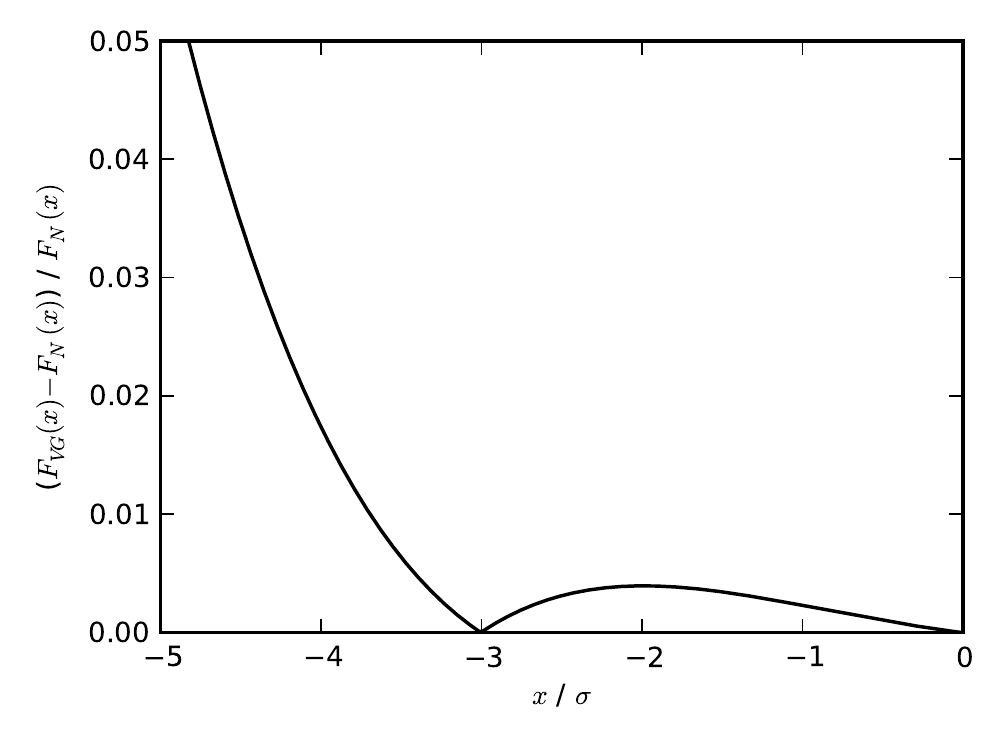}
  \end{center}
  \caption{Absolute relative difference between Variance-Gamma and Normal CDFs. The \textit{x}-axis is rescaled by the Normal standard deviation $\sigma$.}
  \label{VG_Normal_Difference}
\end{figure}

We show the results of this test in Fig.~\ref{VG_Normal_Difference}, plotting the relative difference between the Variance-Gamma\footnote{As in the above examples, for the VG $\theta = 0$ and $k = \frac{1}{2}$.} and Normal CDFs in a percentile region of interest (in Normal terms, between $2\sigma$ and $5\sigma$).
One can see that, already for $n = 50$, the relative deviation at $4\sigma$, corresponding to $P_N(x < -4\sigma) \approx 0.006\%$ is smaller than $2\%$\footnote{The zero value around $x = 3\sigma$ is related to the point where the Normal and VG CDFs intersect.}.
By observing the trend of Fig.~\ref{VG_Convergence} we conclude that, for $n = 250$, the relative differences between the two distributions are immaterial for all calculation purposes.

\section{Application to a test portfolio}\label{Application}

In this section we present empirical results obtained applying our VaR-scaling methodology to a test equity trading portfolio.
To set it up, we chose from the FTSE Index 10 stocks on which we built $\Delta$-hedged positions using at-the-money European call options.
The test portfolio was built in order to mimic the behaviour of realistic positions, although with limited size to be tractable in our numerical experiments.
The (hedged) option positions introduce non-linearity and asymmetry in the P\&L distribution.
The options are struck ATM in order to have maximum time value and convexity, and to rely on most liquid data for their valuation\footnote{In principle the whole volatility surface is quoted, but for most of the single stocks liquid quotes are available only around ATM.}.
The 1-day VaR was calculated using historical simulation (ie, applying historical shocks to the relevant risk factors) as of date 11/02/2014.
A description of the portfolio simulation technique is provided in Appendix~\ref{Appendix_B}.

\begin{figure}[h]
\begin{center}
\includegraphics[width=\textwidth]{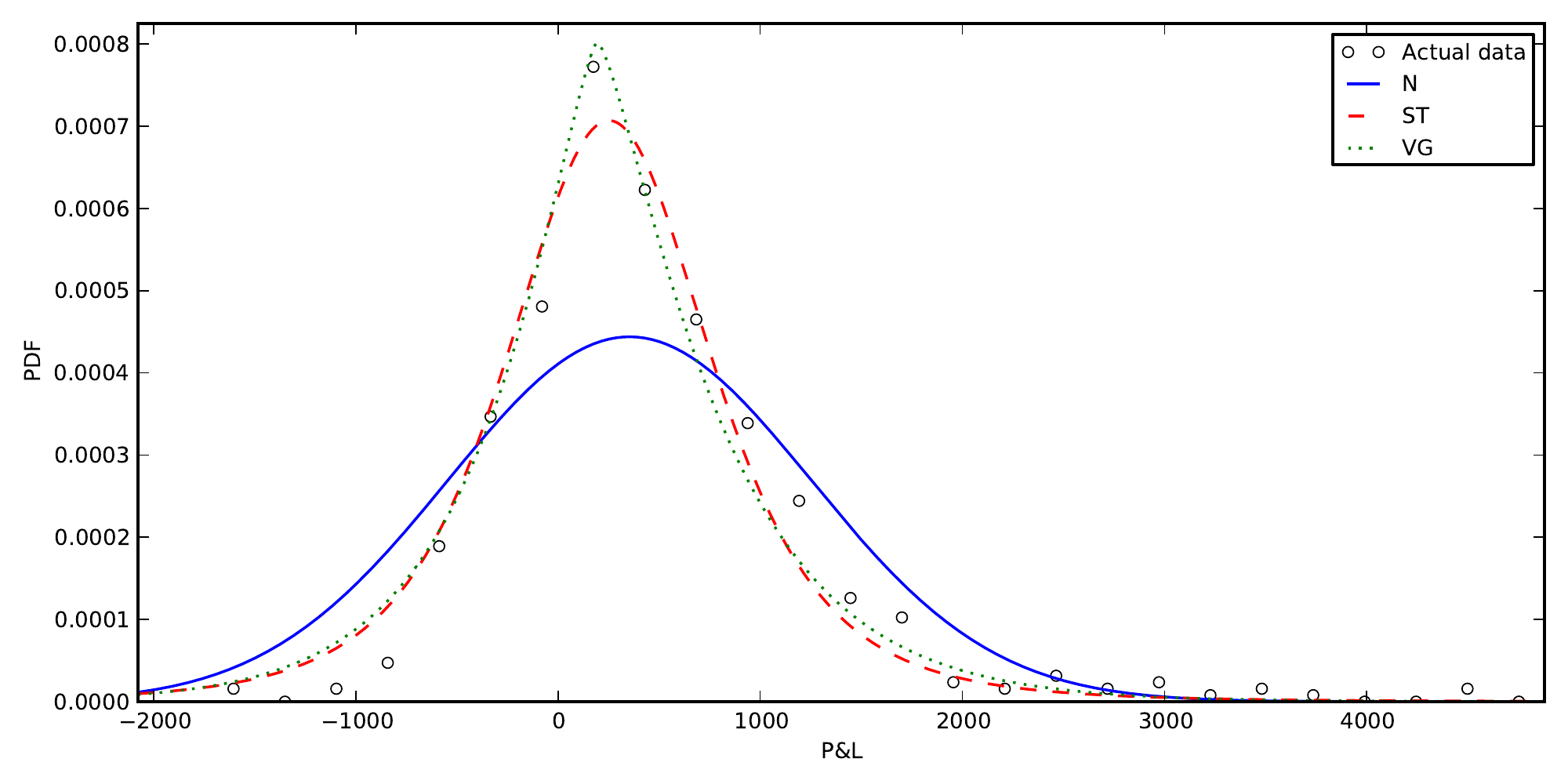}
\end{center}
\caption{Example of a P\&L distribution relative to one of the $N = 10000$ randomly-generated weights combinations. The Normal (N), Student's \textit{t}- (ST) and Variance-Gamma (VG) fits are plotted on top.}
\label{PnL_Example}
\end{figure}

Since it would have been pointless to perform this exercise with an unique combination of weights given to the stocks, we decided to repeat the analysis varying each stock's weight randomly\footnote{A weight extracted from a uniform (0,1) distribution was applied to each stock.}.
In this way, using a high number of repetitions ($N = 10000$), we were able to derive the statistical properties of the portfolio with respect to asset allocation.
To give an example of how the distributions look like, in Fig.~\ref{PnL_Example} we plot the P\&Ls relative to one of the $N$-considered weights combinations.
The fits of Normal, ST and VG distributions are plotted on top of the histogram.

In the following, we comment on the results of our analysis with focus on the Student's \textit{t}-distribution parameter $\nu$ and its statistical behaviour induced by changes in asset allocation.

\subsection{Goodness-of-fit}\label{GoF}
As already observed in the IBM case (Sec.~\ref{PnL_Distributions}), the ST and VG distributions perform better than the Normal in explaining portfolio returns.
To measure the goodness-of-fit, we used the mean ($m$) and standard deviation ($std$) of the MSE over the $N$ samples.
The Normal yields the worst performance, with $m_N \approx 0.07$ and $std_N \approx 0.01$.
The ST distribution achieves better results with $m_{ST} \approx 0.021$ and $std_{ST} \approx 0.003$, while the VG distribution yields $m_{VG} \approx 0.011$ and $std_{VG} \approx 0.003$.

We performed further analyses in order to assess the effect of the skewness in the fit performance.
Specifically, we redid the fit constraining the VG function to be symmetric (setting $\theta = 0$), obtaining new values $m_{VG} = 0.020$ and $std_{VG} = 0.003$, comparable with the ST figures, as expected.
Moreover, we performed an additional test using a non-central Student's \textit{t}-distribution, obtaining MSE results in line with those of the unconstrained VG.

These outcomes are in line with what discussed in Sec.~\ref{PnL_Distributions} regarding the explanatory power of the considered theoretical CDFs, ie, the fit performances of ST and VG distributions are comparable.
Secondary effects are induced by the skewness of the P\&L distribution, and in any case they do not alter significantly the outputs of our analyses performed with the symmetric ST distribution.

\subsection{Student's \textit{t}-distribution parameter $\nu$}\label{ST_Parameters}
\begin{figure}[t]
\begin{center}
\includegraphics[width=\textwidth]{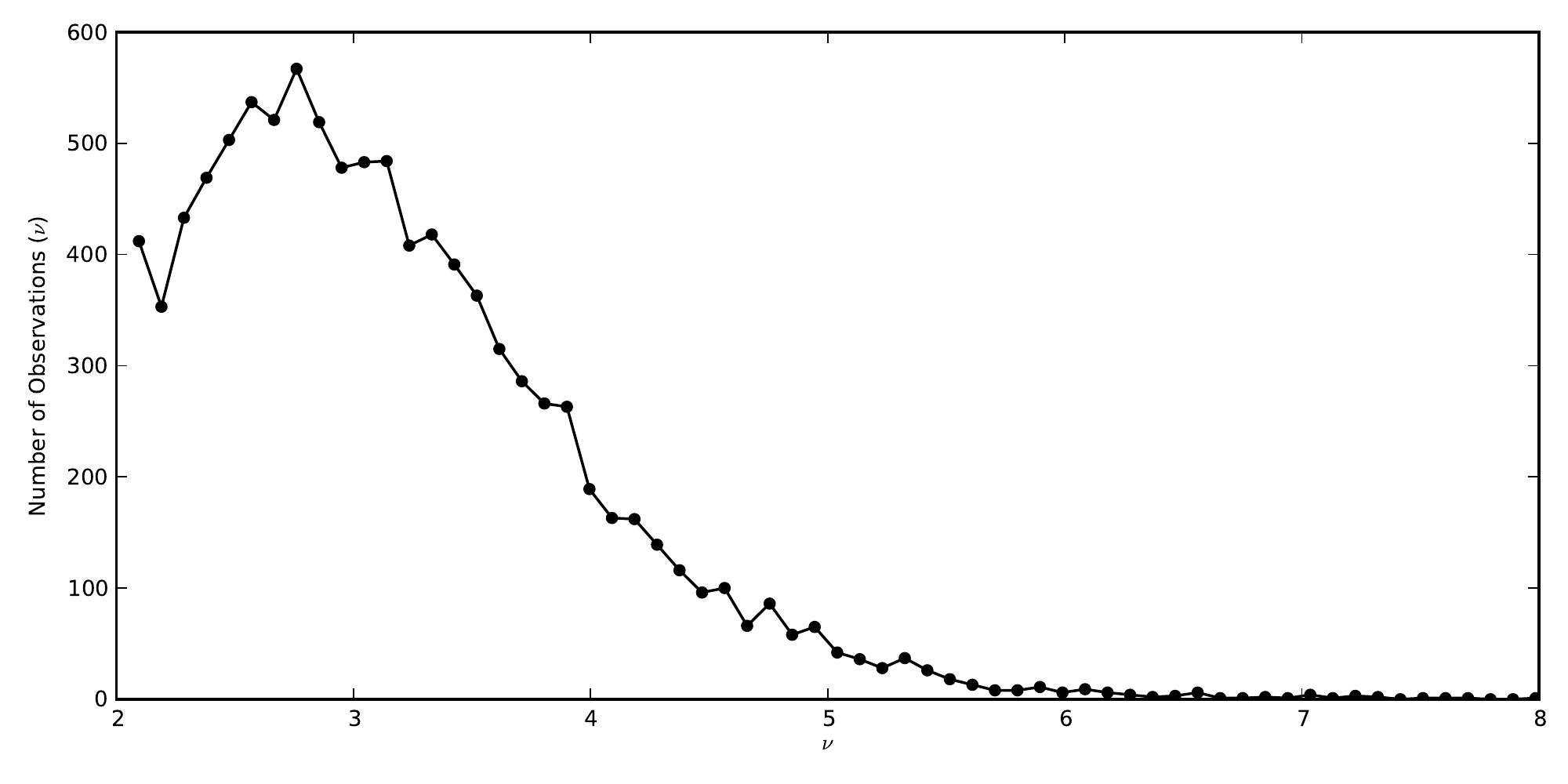}
\end{center}
\caption{Histogram (PDF) of the $N = 10000$ fitted values for parameter $\nu$ of the Student's  \textit{t}-distribution.}
\label{Histogram_Nu}
\end{figure}

In Fig.~\ref{Histogram_Nu} we plot the PDF of the fitted values for parameter $\nu$ of the ST distribution.
If we take into consideration what was derived in Sec.~\ref{Time_Scaling}, we see that the values are generally quite low.
In detail, with mean $\bar{\nu} = 3.13$, standard deviation $\sigma_{\nu} = 0.81$ and approximately 70\% of outcomes lying below the convergence threshold $\nu^* = 3.41$ defined above, we conclude that if it is assumed (or empirically verified) that empirical portfolio returns are best approximated by a ST distribution, it may be not possible to apply the CLT for treating the long-term P\&L distribution as a Normal, since convergence is not achieved at typical confidence levels over a 1-year time horizon.
On the contrary, convolution of the short-term P\&L distribution has to be carried out explicitly.

In the opposite case, ie, when portfolio returns can be modelled by a high-$\nu$ ST distribution, the CLT can be applied, and the terminal distribution can be treated as a Normal.
Given the empirical variance over one step $\sigma^2_{ST}(\Delta t)$, the $T$-convoluted variance of the limiting Normal distribution is simply given by $\sigma^2_N(T) = \sigma^2_{ST}(\Delta t) T$.
\begin{figure}[h]
\begin{center}
\includegraphics[width=\textwidth]{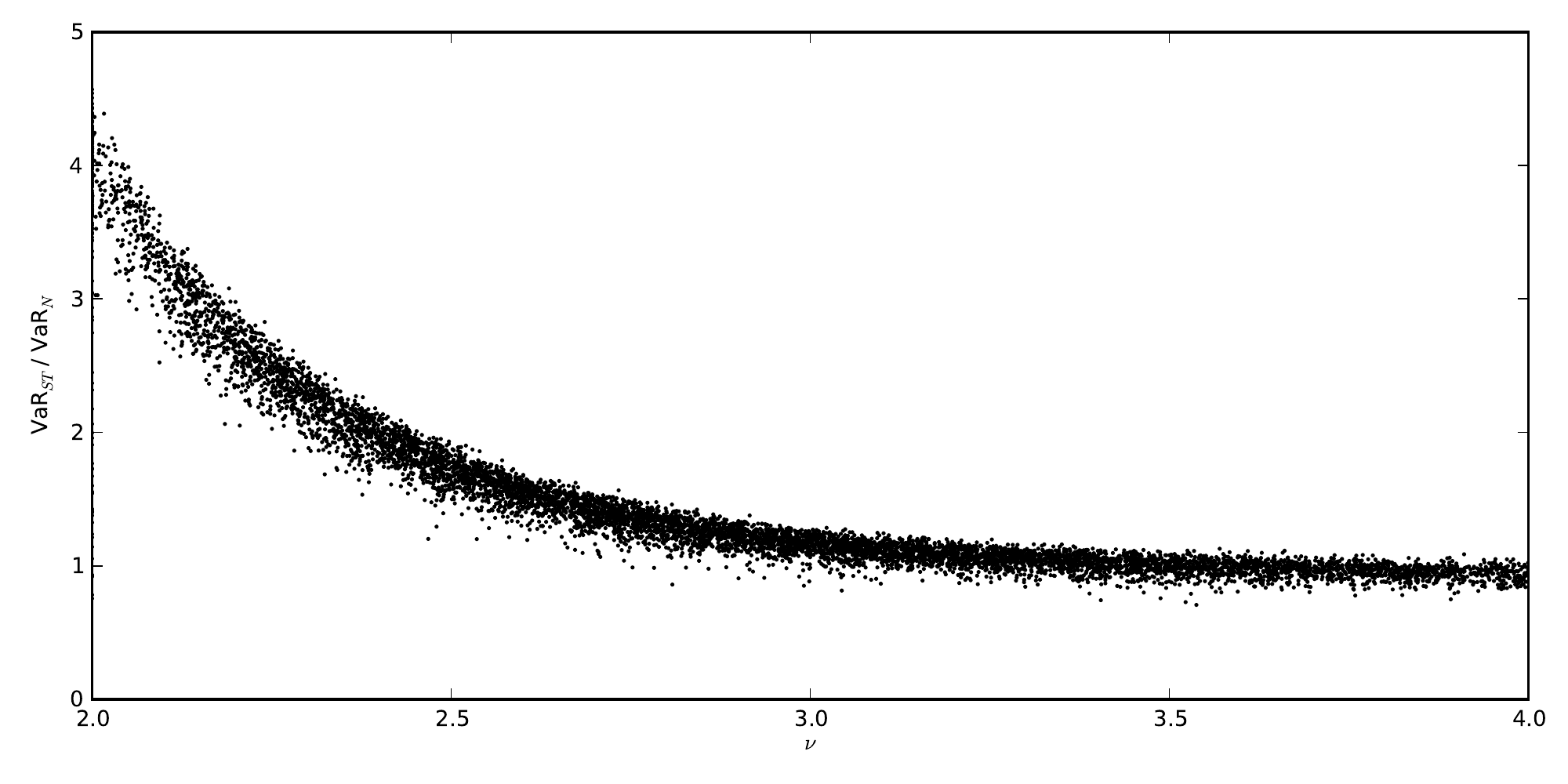}
\end{center}
\caption{Ratio between the 1-year ST VaR and the corresponding Normal VaR for each of the $N = 10000$ portfolio weights as a function of $\nu$.}
\label{VaR_ST_Normal_Ratio}
\end{figure}
\subsection{VaR calculation}\label{VaR_Calculation}

After focusing on the explanatory power of the three distributions we discuss the outcomes of the actual VaR calculations, which we performed in two ways:
\begin{itemize}
\item \textit{Normal VaR}: through application of the CLT, the 1-year P\&L distribution is a Normal with
$$\mu(T) = \mu(\Delta t) T= 0$$
$$\sigma^2(T) = \sigma^2(\Delta t) T$$
where $\mu(\Delta t)$ and $\sigma(\Delta t)$ are the mean and standard deviation of 1-day returns, respectively.
\item \textit{Convoluted VaR}: given the short-term fitted PDF $p_D(x;\cdot)$, convolve it $n = 250$ times to extract the long-term PDF.
This step is executed in different ways depending on the nature of $p_D(x;\cdot)$:
\begin{itemize}
\item If $p_D$ is a VG distribution, the long-term PDF is given by Eq.~\eqref{VG}.
Essentially, as shown in Sec.~\ref{Time_Scaling}, for our time horizon and percentile level the resulting PDF is identical to what would have been obtained applying the \textit{Normal VaR} approach outlined above;
\item If $p_D$ is a ST distribution, the long-term PDF can be estimated numerically by explicitly convolving $p_D$.
We performed this calculation using the Fast Fourier Transform algorithm.
\end{itemize}
\end{itemize}
In Fig.~\ref{VaR_ST_Normal_Ratio} we show the calculated 1-year VaR for different portfolio weights, as a function of the corresponding $\nu$ coefficient of the ST distribution.
Since the portfolio composition varies across the sample, it is more convenient to plot in the \textit{y}-axis the ratio $\frac{\text{VaR}_{ST}}{\text{VaR}_{N}}$ instead of the absolute VaR figures.
As expected, the ST convoluted VaR is generally equal to the Normal VaR in the ST convergence regime ($\nu > \nu^*$)\footnote{The ratio $\frac{\text{VaR}_{ST}}{\text{VaR}_{N}}$ does not converge exactly to 1 (on average). This small bias can be explained considering that the different calibration procedures (moment matching for the Normal and MSE minimization for the ST distribution) may imply a slightly different starting variance for the P\&L distribution.}.
Below the threshold, on the contrary, the ST distribution implies higher VaR levels than the Normal.
Remarkably, the ST VaR grows as high as about four times the Normal VaR as $\nu$ becomes smaller, ie, as the tails of the P\&L distribution become fatter and fatter.

\section{Summary and Conclusion}\label{Summary_Conclusion}

In this paper we derived a generalized VaR-scaling methodology for the calculation of the bank's economic capital.
We considered three PDFs for fitting empirical P\&L distributions and derived accurate scaling rules for each of them.
Starting from the knowledge of a short-term (1-day) distribution, modelling P\&Ls with iid RVs at each time-step we defined the long-term (1-year) distribution by means of convolution.
We also derived the asymptotical properties of the long-term distribution in light of the CLT and investigated, both qualitatively and quantitatively, the convergence conditions under which the CLT can be applied to the considered PDFs for simplifying the EC estimation problem.
In particular, we present an intuitive interpretation of the VaR-scaling problem, allowing to realize whether the implied EC will be over- or underestimated by the usage of the SRTR, given a characterization of the tail behaviour of the short-term P\&L distribution.

The results of this analyses are a range of possible VaR-scaling approaches which depend on
\begin{itemize}
\item the PDF which is chosen as best fit for short-term empirical data,
\item the desired confidence level $1 - \alpha$ and
\item the time horizon $T$ of the Economic Capital calculation.
\end{itemize}
Our numerical examples refer to a typical setting of $1 - \alpha = 99.93\%$ and $T = 1$ year, but our analytical results are valid for any tail confidence level and time horizon.
\begin{figure}[h]

\begin{framed}
\centering
\begin{tikzpicture}[level distance = 1.5cm,level 1/.style={sibling distance=5.5cm},level 2/.style={sibling distance=3.5cm},edge from parent/.style={draw,-latex}]
\tikzstyle{every node}=[rectangle,draw,align=center]

\node (Root){Fit the 1-day P\&L distribution}[edge from parent fork down]
    child
    {
    	node (1){Normal distribution} 
    	child { node(1a) {By definition (SRTR):\\$\sigma^2_{N}(T) = \sigma^2_{N}(\Delta t) T$} }
	}	
	child
    {
    	node (2){Student's \textit{t}-distribution} 
    		child { node (2a){$\nu > \nu^*$}
    			child  { node (2aa){CLT Convergence:\\$\sigma^2_{N}(T) = \sigma^2_{ST}(\Delta t) T$}  } }
    		child { node (2b) {$\nu < \nu^*$}
    			child { node (2ba){\bf Explicit\\\bf convolution} }} 			
	}
	child
	{
		node (3){Variance-Gamma distribution} 
			child { node (3a){CLT Convergence:\\$\sigma^2_{N}(T) = \sigma^2_{VG}(\Delta t) T$}}
	}
	
	;
\end{tikzpicture}
\end{framed}
\caption{Flowchart summarizing the possible VaR-scaling approaches in the calculation of Economic Capital, given a prior knowledge on the short-term (1-day) P\&L distribution.}
\label{Flowchart}
\end{figure}
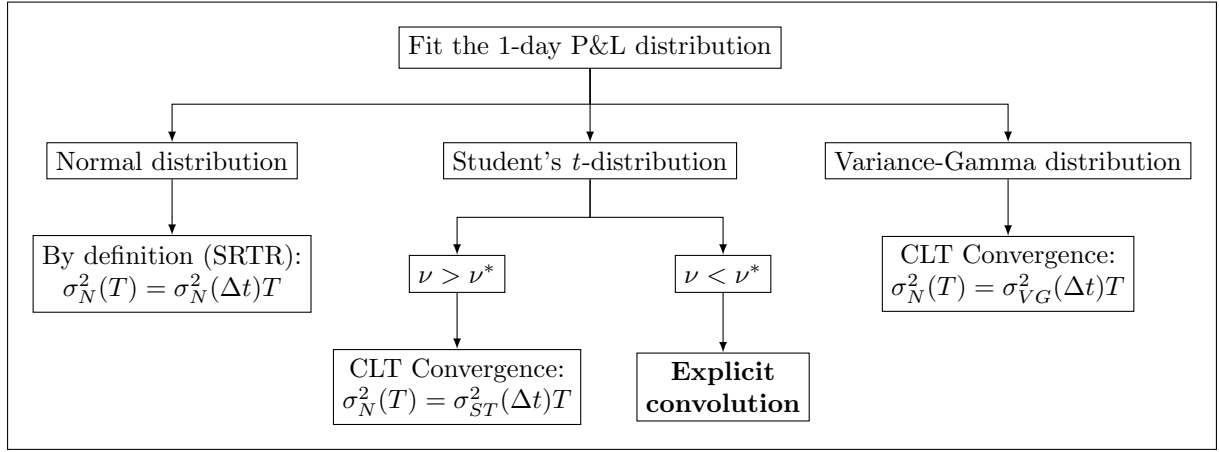
We provide a summary of the possible VaR-scaling techniques outlined in the previous Sections in the flowchart of Fig.~\ref{Flowchart}.
The main discriminant lies in whether the long-term theoretical P\&L distribution has reached Normal convergence or not, given a certain prior knowledge of the short-term P\&L distribution.
When assuming an exponential behaviour of the tails of the PDF (ie, Normal and Variance-Gamma cases), the CLT can be safely applied and the long-term distribution is a Normal whose percentiles can be trivially computed.
When, on the contrary, a power-law decay is assumed (ie, the Student's \textit{t}-distribution case), the CLT can be applied only if the number of degrees of freedom $\nu$ is greater than a critical value $\nu^*$ which depends on the percentile level and the time horizon.
If this is not the case, one needs to compute the $T$-convolution explicitly.
Unfortunately, as outcome of our empirical assessment on a test equity trading portfolio, it appears that this last case could be common in practice, because:
\begin{itemize}
\item We found that the Student's \textit{t}-distribution provides the best fit for empirical P\&L distributions, when the number of observations is high enough to properly identify the tail behaviour;
\item In approximately 70\% of observations (varying asset allocation) the fitted parameter $\nu$ lies below a critical value of $\nu^* = 3.41$.
\end{itemize} 
The choice of the VaR-scaling approach affects substantially the calculation of the Economic Capital.
In particular, the resulting risk-measure can be larger than the estimation obtained using Normal assumptions on the P\&L distribution by up to a factor of four.

Our empirical results on the properties of P\&L distributions, combined with our analytical results on the time scaling of the theoretical PDFs chosen for their modelling, show that the widely-used VaR-scaling technique relying on the application of the SRTR (ie, assuming normality of returns or, at least, an exponential decay in the distribution tails) can lead to a severe underestimation in the bank's long-term risk measure.

\appendix
\section{Student's \textit{t}-distribution approximation accuracy}\label{Appendix_A}
\begin{figure}[t]
\begin{center}
\includegraphics[width=0.5\textwidth]{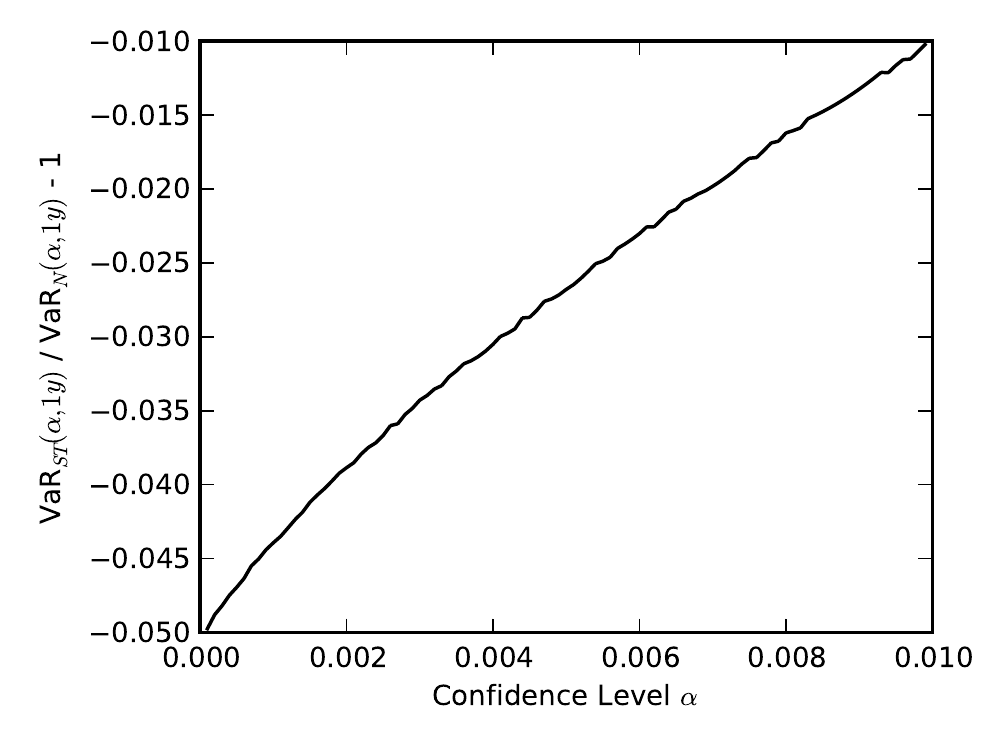}
\end{center}
\caption{Relative differences between Normal and ST VaR considering $\nu^*$ degrees of freedom as a function of the tail probability $\alpha = \text{1 $-$ Condifence Level}$.}
\label{VaR_ST_Normal_Error}
\end{figure}

We consider here the results obtained in Section~\ref{ST_Normal_Convergence} analysing the effects of the approximations therein on the estimation of the EC.
In particular, we show that the relative differences in VaR implied by a $n$-times convoluted Student's \textit{t}-distribution and a Normal with the same variance are negligible when the ST parameter $\nu$ is \textit{equal} to the critical value $\nu^*$, as obtained in Eq.~\eqref{Pcritical}.
In order to estimate quantitatively these differences we performed the following numerical experiment:
\begin{itemize}
\item Fix a confidence level $1 - \alpha$ and estimate $\nu^*$ using Eq.~\eqref{xCritical};
\item Consider a zero-mean ST distribution with $\nu^*$ degrees of freedom and, convoluting it $n=250$ times, obtain a VaR estimation at confidence level $\alpha$;
\item Consider a zero-mean Normal distribution with variance equal to the ST distribution's (ie, equal to $\left(\frac{\sigma^2 \nu^*}{\nu^*-2} \right)n$) and estimate VaR at the same confidence level;
\item Calculate the relative difference between the two VaR estimations.
\end{itemize}
In Fig.~\ref{VaR_ST_Normal_Error} we show the results of our experiment for confidence levels larger than 99.9\%.
Clearly, relative differences in the VaR estimations are quite small for in whole range, the absolute maximum being $5\%$.
Given the results of Section~\ref{Time_Scaling}, we can state that VaR relative differences between a $n$-times convoluted Student's \textit{t}-distribution and a Normal with the same variance are negligible when ST parameter $\nu$ is \textit{greater or equal} to the critical value $\nu^*$.

\section{Portfolio simulation}\label{Appendix_B}

To apply the analytical results of Section \ref{Time_Scaling}, as well as to provide some realistic examples on which to base our discussion, we performed VaR estimations simulating an equity derivatives portfolio.
The portfolio was constructed as follows:
\begin{itemize}
\item Short positions in 10 equities taken from the FTSE Index: Anglo American PLC, Associated British Foods PLC, Antofagasta PLC, ARM Holdings PLC, Aviva PLC, Astrazeneca PLC, BAE Systems PLC, Barclays PLC, British American Tobacco PLC, BG Group PLC;
\item Long $\Delta$-hedged positions in \textit{3-month} ATM european calls on the said equity assets.
This means shorting roughly half the number of underlying shares since for ATM options $\Delta \approx \frac{1}{2}$.
\end{itemize}
The 1-day P\&L distribution is computed using historical simulation, ie, applying historical (daily) shocks in the risk-factors necessary to revaluate the portfolio.
The pricing model used for the options was Black-Scholes with a market-implied volatility.
Some assumptions were made in order to simplify the calculations:
\begin{itemize}
\item A constant dividend yield $q = 3\%$ was assumed for each stock;
\item For discounting equity option flows we used the floating LIBOR 3-month rate $L(0,3m)$;
\item The options were priced assuming no repo margin, and no time lags in equity payments (for calculating forwards) and option payments;
\item The implied volatility used in the pricer for each asset was interpolated bi-linearly using the current surface (deduced from market option prices). A unique relative shock (equal to the relative historical shock in the ATM \textit{3-month} volatility) was applied to the whole surface.
\end{itemize}
Although the above assumptions introduce (under a pure pricing point-of-view) some slight inaccuracy in the revaluation of the portfolio, the main goal of this simulation was to highlight the non-normality of empirical returns in (equity) derivatives portfolios.
The non-normality of the P\&L distribution is ultimately originated by 1) non-normality in equity historical returns, 2) non-linearity in portfolio positions (due to options) and 3) the volatility smile kept into account in the revaluation of options prices.

\printbibliography

\end{document}